\documentclass[12pt]{article}
\usepackage{jheppub}
\usepackage{tikz}
\usetikzlibrary{positioning}
\usetikzlibrary{calc}
\usepackage{physics}
\usepackage{simplewick}
\usepackage{simpler-wick}

\title{Twisted holography without conformal symmetry}
\author[a,b]{Kasia Budzik,}
\author[a]{Davide Gaiotto}

\affiliation[a]{Perimeter Institute for Theoretical Physics, Waterloo, ON N2L 2Y5, Canada}
\affiliation[b]{Department of Physics \& Astronomy, University of Waterloo, Waterloo, ON N2L 3G1, Canada}

\emailAdd{kbudzik@perimeterinstitute.ca}
\emailAdd{dgaiotto@perimeterinstitute.ca}

\abstract{We discuss the notion of translation-invariant vacua for 2d chiral algebras and relate it to the notion of the associated variety. The two-dimensional chiral algebra associated to four-dimensional ${\cal N}=4$ $U(N)$ SYM has a conjectural holographic dual involving the B-model topological string theory. 
We study the effect of non-zero vacuum expectation values on the chiral algebra correlation functions and derive a holographic dual Calabi-Yau geometry. We test our proposal 
by a large $N$ analysis of correlation functions of determinant operators, whose saddles can be matched with semi-classical configurations of ``Giant Graviton" D-branes in the bulk.}


\begin{document}
\maketitle

\section{Introduction and Conclusions}
Any ${\cal N}=2$ superconformal field theory in 4d contains a 2d chiral algebra\footnote{We use the terms 2d chiral algebra and vertex algebra interchangeably throughout the paper. Conceptually, a 2d chiral algebra encodes the properties of holomorphic local operators supported on a two-dimensional locus in some quantum field theory. Vertex algebras provide a mathematical formalization of 2d chiral algebra, essentially via Kaluza-Klein reduction on a circle.} subsector, protected by a linear combination ``$Q+S$'' of super-charges and super-conformal charges of the ambient theory \cite{Beem:2013sza}. Many such SCFTs are also equipped with a non-trivial moduli space of super-Poincar\'e-invariant vacua. The space of vacua may include various ``branches'', distinguished by the type of BPS operators which acquire vevs. The Higgs branch of vacua can be characterized as leaving unbroken the $U(1)_r$ R-symmetry of the SCFT. According to a conjecture of \cite{Beem:2017ooy}, one can recover the Higgs branch of the space of 4d vacua as the \textit{associated variety} of the 2d chiral algebra\footnote{See also \cite{Song:2017oew}, and the reviews \cite{Arakawa:2017aon, Arakawa:2017fdq}.}, an algebraic tool introduced originally to study the characters of a vertex algebra \cite{2010arXiv1004.1492A}.


This conjecture is surprising: Higgs branch vevs break the 4d superconformal group to the super-Poincar\'e group and in particular break the ``$Q+S$'' symmetry protecting the chiral algebra subsector. The tension can be resolved by considering an alternative definition of the chiral algebra subsector based on a B-type $\Omega$-deformation of the physical theory \cite{Jeong:2019pzg, Butson:2020mmu, Oh:2019bgz}: the B-type $\Omega$-deformation setup only requires unbroken $U(1)_r$ R-symmetry\footnote{This should be contrasted to the $\Omega$-deformation defined in \cite{Nekrasov:2002qd}, which is instead applicable to any ${\cal N}=2$ SQFT.} and thus appears compatible with Higgs branch vevs.\footnote{We leave a proof of this fact to future work. We can find support for this assumption in the main example in this paper: the chiral algebra of $U(N)$ ${\cal N}=4$ SYM. It we identify the theory as the low-energy world-volume theory on D3 branes, the B-type $\Omega$-deformation is induced by a string theory $\Omega$-background, which reduces the D3 branes to certain B-branes in the B-model topological string. These B-branes have a moduli space of vacua parameterized by the Higgs branch of the physical theory.} This leads to the obvious question: how does a Higgs branch vev affect the chiral algebra correlation functions computed in the $\Omega$-deformed theory?

In order to answer this question, we introduce the notion of translation-invariant vacuum for a 2d chiral algebra. A simple analysis, presented in Section \ref{sec:three} of this note, leads to a transparent physical interpretation of the associated variety: it coincides with the moduli space of translation-invariant vacua of the 2d chiral algebra. The conjecture of \cite{Beem:2017ooy} thus identifies the Higgs branch vacua of the 4d ${\cal N}=2$ SCFT with the vacua of the corresponding 2d chiral algebra. In the $\Omega$-deformation setup, the identification maps the expectation values of chiral algebra operators to the expectation values of the corresponding operators in the 4d theory.

Correlation functions of a 2d chiral algebra in a non-trivial vacuum are a novel observable, which may be of mathematical interest.\footnote{It should be also be possible to insert non-trivial modules for the chiral algebra at points in the plane and define conformal blocks in a non-trivial vacuum.} In the rest of the paper, we restrict our attention to ${\cal N}=2$ SCFTs which have a Type IIB holographic dual \cite{Maldacena:1997re, Aharony:1999ti}. Most of our calculations will be done in the chiral algebra of ${\cal N}=4$ $U(N)$ gauge theory, but can be extended to other examples with some extra work.

One of the simplest variations of Maldacena duality \cite{Maldacena:1997re} involves precisely correlation functions of $SU(N)$ ${\cal N}=4$ SYM computed in non-trivial super-Poincar\'e-invariant vacua, i.e. the Coulomb branch,\footnote{The $\mathcal{N}=4$ Coulomb branch is $(\mathbb{R}^6)^N/S_N$. It includes the $\mathcal{N}=2$ Coulomb and Higgs branches $\mathbb{C}^N/S_N$ and $(\mathbb{C}^2)^N/S_N$, parametrized respectively by vevs of two and four of the six scalars $\vec\Phi$.} in flat space \cite{Douglas:1998tk, Bilal:1998ck, Wu:1998hx, Tseytlin:1998cq, Kraus:1998hv}, where the six adjoint scalar fields $\vec\Phi$ receive diagonal vevs, with eigenvalues $\vec y_i$ of multiplicities $N_i \equiv \alpha_i N$, $i=1,\dots, n$. 

The dual IIB supergravity solutions are obtained from a near-horizon limit of multi-center half-BPS D3 brane solutions:
\begin{equation}
\dd s^2 = H(\vec y)^{-\frac12} \dd x^2 + H(\vec y)^{\frac12} \dd \vec y \, ^2 \, ,
\end{equation} 
where $x$ are the four directions parallel to the D3 branes and $\vec y$ the six directions transverse to the D3 branes. The solutions 
involve a harmonic function $H(\vec y)$ in $\mathbb{R}^6$:
\begin{equation}
H(\vec y) = L^4 \sum_{i=1}^n \frac{\alpha_i}{|\vec y - \vec y_i|^4} \, ,
\end{equation}
with $\vec y_i$ being the transverse positions of the D3 brane stacks, each with a fraction $\alpha_i$ of the total number $N$ of D3 branes.\footnote{The solution for D3 branes in flat space has an extra ``1'' constant term in $H$, which drops out in the near-horizon limit.}

The $\mathbb{R}^4$ holographic boundary\footnote{As these vacua are not conformally invariant, correlation functions do not naturally extend to the four-sphere.} lies at $\vec y \to \infty$ . The asymptotic deviation of $H(\vec y)$ from the $AdS_5 \times S^5$ reference value $L^4 |\vec y|^{-4}$ encodes the vacuum expectation values of local operators, according to the standard holographic dictionary.

The main objective of this note is to study the analogous setting for the twisted holography correspondence of \cite{Costello:2018zrm}, which is expected to capture a protected subsector of Maldacena's duality. This correspondence relates the 2d chiral algebra subsector of 4d ${\cal N}=4$ SYM \cite{Beem:2013sza} to the B-model topological string theory \cite{Bershadsky:1993cx,Bershadsky:1993ta} on appropriate 3d (complex) Calabi Yau geometries \cite{Gopakumar:1998ki} endowed with a holographic boundary.

The Calabi-Yau geometry dual to the 2d chiral algebra in the standard conformally-invariant vacuum is the deformed conifold, i.e. $SL(2,\mathbb{C})$. We follow closely the derivation of \cite{Costello:2018zrm} and present in section \ref{sec:one} a family of candidate 3d Calabi-Yau geometries dual to the correlation functions of the 2d chiral algebra in a non-trivial vacuum. 

As a test of our proposal, we study correlation functions involving determinant operators in the chiral algebra. In the full physical theory, the insertion of such determinant operators in the boundary theory is dual to ``Giant Graviton'' D3 branes approaching a point of the holographic boundary of $\text{AdS}_5\times S^5$. In twisted holography, the D-branes wrap 1-dimensional complex curves in the Calabi-Yau geometry \cite{Budzik:2021fyh}. In either case, 
the large $N$ saddles of correlation functions of multiple determinants should be dual to semi-classical D-brane configurations in the bulk \cite{Jiang:2019xdz, Budzik:2021fyh, Jiang:2019zig, Yang:2021kot} (and references therein).

In the standard conformally-invariant vacuum, a spectral curve construction maps the large $N$ saddles of chiral algebra correlation functions to explicit complex curves in $SL(2,\mathbb{C})$ \cite{Budzik:2021fyh}. In section \ref{sec:two} we will use a similar but somewhat more intricate construction to extend the match to the saddles which appear for non-trivial vacua.

\section{Translation-invariant vacua and associated varieties}\label{sec:three}
In this section we introduce the notion of translation-invariant vacua for a 2d chiral algebra and propose it coincides with the associated variety of the chiral algebra.

A translation-invariant vacuum $\mathcal{V}$ is a collection of translation-invariant correlation functions on the plane which satisfy OPE and have the cluster property: the correlation functions factorize in the limit where a subset of the local operators is 
far from the rest. In particular, correlation functions in the vacuum $\cal V$ should have a finite limit 
\begin{equation}
\left \langle O_0(z) \prod_a O_a(z_a) \right \rangle_{\cal V} \to \langle O_0 \rangle_{\cal V}\, \left \langle  \prod_a O_a(z_a) \right \rangle_{\cal V}, \quad \quad z \to \infty
\end{equation}
when one operator is brought to infinity. 

We can use a standard strategy to determine the dependence of such a correlation function on the position $z$ of one operator: 
the OPE relations determine its poles as a function of correlation functions of fewer operators and factorization controls the behaviour at infinity. 
Every correlation function can be reconstructed recursively in this manner, given the one-point functions $\langle O_a \rangle_{\cal V}$. 

The full OPE relations contain more information than the singular parts. The non-singular part of the OPE places constraints on 
the one-point functions $\langle O_a \rangle_{\cal V}$. For example, consider a two-point function 
\begin{equation}
\langle O_a(z) O_b(0) \rangle_{\cal V} = \langle O_a \rangle_{\cal V}\langle O_b \rangle_{\cal V} + \sum_{n\geq 0} z^{-n-1} \langle [O_{a;n} O_b] \rangle_{\cal V} \, ,
\end{equation}
reconstructed from the behaviour at large $z$ and the singular part of the OPE.\footnote{We use the mathematical conventions here, so that $n=0,1,\dots$ correspond to the singular part and $n=-1,-2,\dots$ correspond to the finite part of the OPE.} If we compare this to the full OPE expansion,
\begin{equation}
\langle O_a(z) O_b(0) \rangle_{\cal V} =\sum_{n\in\mathbb{Z}} z^{-n-1} \langle [O_{a;n} O_b] \rangle_{\cal V} \, , \label{eq:OPE}
\end{equation}
we deduce the one-point functions of all operators which appear in the non-singular part of the OPE: \footnote{Notice that $[O_{a;-n} O_b]$ for positive $n$ is just the regularization of a $\partial^{n-1} O_a \,O_b$ composite operator. The relations below just tell us that the regularization does not affect the factorization of vevs. }
\begin{align} \label{eq:commute}
&\langle [O_{a;-n} O_b] \rangle_{\cal V}  =0 \qquad \qquad  \qquad n \geq 2 \cr
&\langle [O_{a;-1} O_b] \rangle_{\cal V}  =\langle O_a\rangle_{\cal V}\langle O_b \rangle_{\cal V}  \, .
\end{align}

The space of operators of the form $[O_{a;n} O_b]$, $n\leq -2$, forms a very nice subspace $C_2(V)$ of the vertex algebra $V$.\footnote{If we take $O_b$ to be the identity, we find that derivatives of any operator belong to $C_2(V)$.} The quotient $R_V=V/C_2(V)$ equipped with the (commutative) product $O_a\cdot O_b \equiv [O_{a;-1} O_b]$ ($\textrm{mod}\, C_2(V)$) is called the Zhu's $C_2$-algebra of the VOA \cite{zhu1990vertex}. 

The relations (\ref{eq:commute}) are equivalent to the statement that the 1-point functions $\langle O_a \rangle_{\cal V}$ define an algebra map from $R_V$ to the complex numbers. 
Essentially by definition,\footnote{The maximal spectrum $\textrm{mSpec} \, R$ of a commutative $\mathbb{C}$-algebra $R$ is defined as the set of its maximal ideals. Quotiening $R$ by a maximal ideal results in the field $\mathbb{C}$. Conversely, kernel of any algebra map from $R$ to $\mathbb{C}$ is a maximal ideal.} this is the same as a point in the associated variety of the VOA, which is defined as the maximal spectrum of the $C_2$-algebra \cite{2010arXiv1004.1492A}:
\begin{align}
\mathcal{X}_V = \textrm{mSpec} \, R_V \, .
\end{align}
Conversely, any algebra map from $R_V$ to the complex numbers gives us a collection of 1-point functions, with the property that the 2-point functions derived from those 
via Ward identities satisfy cluster decomposition and are compatible with the full OPE. 

We expect this property to be sufficient to guarantee that all $n$-point-functions also satisfy cluster decomposition and are compatible with the full OPE expansion. It would be nice to prove this fact. With that assumption, we find that the space of vacua for $V$ coincides with $\mathcal{X}_V$. 

The $C_2$-algebra is also equipped with a Poisson bracket $\{ O_a, O_b \} \equiv [O_{a;0} O_b]$. The corresponding Hamiltonian flows $\{ O_a, \cdot \}$ have a natural physical interpretation: they describe the infinitesimal deformation of the vacuum induced by integrating $O_a$ on a very large circle.\footnote{The insertion of such contour integral indeed preserves translation symmetry and the cluster property: the contour can be translated and can also be deformed to separately encircle two collections of well-separated local operators.} If $O\equiv J$ is a dimension $1$ current, the zero-mode of $J$ defines a symmetry of $V$ and the Poisson bracket $\{ J, \cdot \}$ is the action of the same symmetry on $R_V$. Equivalently, the image of $J$ in $R_V$ is the moment map for the symmetry associated to $J$.

\subsection{Gauged $\beta\gamma$ systems}
\label{subsec:betagamma}

As an example, we consider gauged $\beta\gamma$ systems, which arise as chiral algebras of ${\mathcal{N}=2}$ Lagrangian gauge theories. We identify their associated varieties and translation-invariant vacua as well as match them with the Higgs branches of the corresponding 4d theories.

First, consider a particularly simple example of associated variety, which occurs for $\beta \gamma$ system of free symplectic bosons $\mathrm{Sb}$, with OPE
\begin{equation}
	Y(z) X(0) \sim \frac{1}{z} \, .
\end{equation}
It is easy to see that any operator which contains derivatives of the elementary fields belongs to $C_2(\mathrm{Sb})$. Then $R_{\mathrm{Sb}}$ consists of polynomials in two variables $x=[X]$ and $y=[Y]$ ($\textrm{mod } C_2(\textrm{Sb})$) and the associated variety is $\mathcal{X}_{\mathrm{Sb}}=\mathbb{C}^2$, with the standard Poisson bracket $\{y,x\}=1$. Correspondingly, the translation-invariant vacua of $\mathrm{Sb}$ are labelled by the vevs 
\begin{equation}
x = \langle X(0) \rangle_{x,y} \, , \qquad \qquad y = \langle Y(0) \rangle_{x,y} \, .
\end{equation} 
The associated 4d $\mathcal{N}=2$ theory is a single free hypermultiplet, whose Higgs branch is indeed $\mathbb{C}^2$. The generalization to multiple copies of the symplectic boson chiral algebra is straightforward.

There exists a universal prescription for the associated variety of a gauged chiral algebra in terms of the associated variety of the origical chiral algebra. Consider a 2d chiral algebra $V$ equipped with a Kac-Moody symmetry $G$ at level $- 2 h$, with $h$ being the dual Coxeter number of $G$. Such a Kac-Moody symmetry can be gauged to produce a new 2d chiral algebra, which we denote $V/\!\!/G$. Concretely, one adds a $bc$ ghost system for $G$ and takes cohomology with respect to a standard BRST charge. The associated variety for  $V/\!\!/G$ is known to coincide with the complex symplectic quotient\footnote{Recall that the complex symplectic quotient $\mathcal{X} /\!\!/G$ of a variety $\mathcal{X}$ with Hamiltonian G-action is defined by quotiening the vanishing locus for the moment maps  by (the complexification of) $G$. The moment maps are the Hamiltonians for the $G$ action.} $\mathcal{X}_V/\!\!/G$ of the associated variety of $V$ \cite{Arakawa:2010ni, Arakawa:2018egx}. 

The proof in the last reference essentially shows that the operation of taking Zhu's $C_2$-algebra $R_V$ commutes with BRST reduction, in the sense that 
$R_{V/\!\!/G}$ is obtained from $R_V$ by adding the ghost representatives $[b]$ and $[c]$ and defining a BRST charge as the Poisson bracket with the representative of the BRST current $\{J_{\text{BRST}}, \cdot \}$.
The BRST reduction of $R_V$ is a derived description of the complex symplectic quotient of the corresponding associated variety: it makes the moment maps BRST exact and imposes $G$ invariance.

More physically, this theorem tells us that a vacuum for $V/\!\!/G$ is the same as a BRST-invariant vacuum for the combination of $V$ and the ghosts. The BRST variation of the vacuum is computed as an integral of the BRST current on a large circle, which is the same as the Poisson bracket with the current representative in $R_V$. This statement is fully compatible with the conjectural relation between associated varieties and Higgs branches. Indeed, consider a 4d  ${\cal N}=2$ SCFT with global symmetry $G$. As long as the $U(1)_r$ symmetry does not become anomalous, gauging $G$ (in an ${\cal N}=2$ sense) results in a new 4d  ${\cal N}=2$ SCFT. The Higgs branches of the two theories are related by a complex symplectic quotient and the 2d chiral algebras of the two theories are related by gauging as well. Incidentally, this reasoning also proves the conjecture for all Lagrangian theories, which give rise to gauged $\beta \gamma$ systems \cite{Beem:2013sza, Beem:2017ooy}.

\subsubsection{Chiral algebra of $\mathcal{N}=4$ SYM}
In the case of ${\cal N}=4$ SYM, the $\beta \gamma$ system is a pair symplectic bosons $X$, $Y$ in the adjoint representation of $U(N)$ \cite{Beem:2013sza}. The OPE is
\begin{equation}
Z^a_b(u;z)Z^c_d(v;w) \sim \frac{1}{N} \frac{u-v}{z-w} \delta^a_d \delta^c_b \, ,
\end{equation}
where $Z(u;z)$ is the linear combination
\begin{equation}
Z(u;z) \equiv X(z) + u Y(z) \, .
\end{equation}
The purpose of the auxiliary variable $u$ is to write expressions which are covariant under the $SL(2)_R$ symmetry\footnote{This is a subgroup of the R-symmetry of the 4d theory.} 
rotating $X$ and $Y$ into each other, which acts as fractional linear transformations on $u$. 

The moment map is $[X,Y]$ and we can select a vacuum where the $X$ and $Y$ receive vevs that are any commuting diagonal matrices, with eigenvalues $(x_i,y_i)$ appearing with multiplicity $N_i$ (by definition, $\sum_i N_i =N$). The vev eigenvalues represent the positions of stacks of $N_i$ D-branes in the transverse $\mathbb{C}^2$. 

The simplest set of single-trace BRST-closed local operators, the \textit{$A$-tower}, consist of the individual terms of the expansion of 
\begin{equation}
A_n(u;z) = N\Tr Z(u;z)^n
\end{equation}
in powers of $u$, i.e.
\begin{equation}
A_{n;s}(z) = N\text{STr}\, X^{n-s} Y^s (z)\, ,
\end{equation}
where $\text{STr}\, X^{n-s} Y^s$ is the symmetrized trace. These form an irreducible representation of $SL(2)_R$ of dimension $(n+1)$.

In a vacuum parametrized by eigenvalues $(x_i,y_i)$ of $X$ and $Y$ of multiplicity $N_i$, they receive vacuum expectation values
\begin{equation}
\langle \mathrm{STr} X^n Y^m \rangle =  \sum_i N_i x_i^n y_i^m \, ,
\end{equation}
which we will give a holographic interpretation in the next section.

\section{Coulomb branch geometry} \label{sec:one}

In this section, we generalize the analysis of section 4 of \cite{Costello:2018zrm} and compute the backreaction of a collection of parallel \textit{non-coincident} D-branes in the B-model/BCOV theory. We identify the backreacted geometry as dual to the chiral algebra of $\mathcal{N}=4$ SYM in the non-trivial vacua studied above.

Consider B-model branes wrapping parallel $\mathbb{C}$'s in $\mathbb{C}^3$. We use coordinates $(x,y,z)\in\mathbb{C}^3$ so that $N_i=\alpha_i N$ branes wrap $\mathbb{C}$ defined by equations $x=x_i$, $y=y_i$.
The back-reaction is described by a Beltrami differential \footnote{For convenience, we set the topological string coupling to $N^{-1}$.}
\begin{equation}
\beta = \sum_{i=1}^n \alpha_i \frac{(\bar x- \bar x_i)\dd \bar y- (\bar y- \bar y_i)\dd \bar x}{(|x-x_i|^2+|y-y_i|^2)^2} \partial_{z} \, ,
\end{equation}
which deforms the complex structure of $\mathbb{C}^3\setminus\bigcup_i\{x=x_i,y=y_i\}$.

We can also give a \v{C}ech description of this Beltrami differential. The description involves $2^n$ patches of $\mathbb{C}^3$, in which, for each $i=1,\dots,n$, either 
$x-x_i$ or $y - y_i$ is non-zero. We denote a patch by an index $I$ and denote as $I_x$ the collection of $i$'s for which $x-x_i$ is non-zero and as $I_y$ the collection of $i$'s for which $y-y_i$ is non-zero. We can trivialize the Beltrami differential in each patch by a gauge transformation, giving us a new holomorphic local coordinate $z_I$.\footnote{In a patch $I$, $\beta$ can be written as $\beta=\bar\partial\gamma_I$, where
\begin{equation}
\gamma_I=\sum_{i\in I_x} \qty(\frac{\bar y-\bar y_i}{x-x_i})\frac{\alpha_i}{|x-x_i|^2+|y-y_i|^2}-\sum_{i\in I_y}\qty(\frac{\bar x-\bar x_i}{y-y_i})\frac{\alpha_i}{|x-x_i|^2+|y-y_i|^2} \, .
\end{equation}
The new holomorphic local coordinate is then $z_I=z-\gamma_I$.} The coordinates $x$ and $y$ are holomorphic and coincide in all patches. 


On the intersection of two patches $I$ and $I'$ which differ by the $i$-th choice only, where $I_y$ and $I'_x$ include $i$, we have a coordinate transformation 
\begin{equation}
z_I = z_{I'} + \frac{\alpha_i}{(x-x_i)(y - y_i)} \, . \label{eq:IIprime}
\end{equation}
More generally, $z_I - z_{I'}$ is a sum of terms, with positive sign for each $i$ included both in $I_y$ and $I'_x$ or with negative sign for each $i$ included both in $I_x$ and $I'_y$.

We can equivalently use coordinates  
\begin{equation}
w_I = z_I \left[\prod_{i \in I_x} (x-x_i)\right] \left[ \prod_{i \in I_y} (y-y_i) \right] ,
\end{equation}
which extend to globally defined functions on the whole geometry and satisfy relations
\begin{equation}
w_I (x-x_i) - w_{I'} (y-y_i) = \alpha_i \left[\prod_{j \in I_x} (x-x_j)\right] \left[ \prod_{j \in I'_y} (y-y_j) \right] \, , \label{eq:coordw}
\end{equation}
whenever $I$ and $I'$ differ at the $i$'th entry only, where $I_y$ and $I'_x$ include $i$. 

The holographic boundary is at $x,y\rightarrow\infty$. The coordinate $z \sim z_I$ parameterizes the boundary and is identified with the 
holomorphic coordinate on the 2d chiral algebra plane.\footnote{As we mentioned in a previous footnote, correlation functions in this non-conformal setup are only defined on a plane. We do not add a point at infinity to extend the holographic boundary to  $\mathbb{CP}^1$, as was done in \cite{Costello:2018zrm}.} Holographic boundary conditions were formulated in a holomorphic language in  \cite{Costello:2018zrm}. 

The holomorphic formulation of the boundary conditions makes use of an ``internal'' $\mathbb{CP}^1$ defined by the ratio $x/y$ as $x$ and $y$ are sent to $\infty$. In our chosen gauge, the intricacies of the deformed geometry appear asymptotically in a neighbourhood of the poles of the internal $\mathbb{CP}^1$: directions with generic $x/y$ belong to the intersection of all patches $I$. Near the North and South poles of $\mathbb{CP}^1$, we can assume we are sitting respectively in either of the two ``extremal'' patches, $I_0$ where $y - y_i$ are all non-zero or $I_\infty$ where $x - x_i$ are all non-zero. The coordinate transformation between the two patches is 
\begin{equation}
z_0 - z_\infty = \sum_{i=1}^n \frac{\alpha_i}{(x-x_i)(y-y_i)} =\sum_{k\geq 0}\sum_{l\geq 0}\frac{1}{x^{k+1} y^{l+1}} \sum_{i=1}^n \alpha_i x_i^k y_i^l \, . \label{eq:dualgeo}
\end{equation}
Each individual $\frac{1}{x^{k+1} y^{l+1}}$ term after the first describes a deformation of the standard $SL(2,\mathbb{C})$ geometry\footnote{Using coordinates $w_0=z_0y$, $w_\infty=z_\infty x$, we get the familiar $SL(2,\mathbb{C})$ relation: $x w_0 - y w_\infty = 1$.}
\begin{equation}
z_0 - z_\infty = \frac{1}{x y} 
\end{equation}
obtained by backreation of a coincident stack of branes at $x=y=0$. 

The deformation decays as we approach the boundary and represents holographically the vevs 
\begin{equation}
\langle \mathrm{STr}\, X^k Y^l \rangle =  \sum_{i=1}^n N_i x_i^k y_i^l
\end{equation}
of the corresponding chiral algebra local operators in a vacuum where $X$ and $Y$ go to commuting expectation values at $\infty$, with eigenvalues $(x_i,y_i)$ of multiplicity $N_i$. 

This confirms our identification of the geometry as a description of the chiral algebra correlation functions in a non-trivial vacuum. 

\subsection{Restoring $SL(2)_R$ invariance}
The original $\mathbb{C}^3$ geometry has an $SL(2)_R$ symmetry rotating the $x$ and $y$ coordinates, which was broken by our choice of trivializations: the coordinates we use come from the trivializations
\begin{equation}
\frac{\bar x d \bar y- \bar yd \bar x}{(|x|^2+|y|^2)^2} = \bar \partial \left( \frac{\bar y}{x (|x|^2+|y|^2)}\right)= \bar \partial \left( -\frac{\bar x}{y (|x|^2+|y|^2)}\right) \, ,
\end{equation}
which privilege respectively the $x$ or $y$ coordinates.

A more general trivialization would be 
\begin{equation}
\frac{\bar x d \bar y- \bar yd \bar x}{(|x|^2+|y|^2)^2} = \bar \partial \left( \frac{\bar y- v \bar x}{(x+ v y) (|x|^2+|y|^2)}\right) \, ,
\end{equation}
which interpolates between the two as we vary the parameter $v$. 

Such a trivialization gives us another family $z_{v^{-1}}$ of local coordinates on the deformed geometry (in a patch avoiding the $x+v y = x_i + v y_i$ loci):
\begin{equation}
z_{v^{-1}} = z_0 + \sum_{i=1}^n \frac{\alpha_i}{(x-x_i+ v (y-y_i) )(y - y_i)}= z_\infty- \sum_{i=1}^n \frac{\alpha_i v}{(x-x_i)(x-x_i+ v (y-y_i) )} \, .
\end{equation}
Different members of the family are related as 
\begin{equation}
z_{u^{-1}} -z_{v^{-1}} =  \sum_{i=1}^n \frac{\alpha_i (v-u)}{(x-x_i+ u (y-y_i) )(x-x_i+ v (y-y_i) )} \, .
\end{equation}
These coordinates will be useful below. 

\section{Determinant correlation functions in a Coulomb background}\label{sec:two}

Recall, the chiral algebra subsector \cite{Beem:2013sza} of $\mathcal{N}=4$ SYM is a gauged system of symplectic bosons $X,Y$ in the adjoint representation of $U(N)$. 

We will now compute correlation functions of determinant operators in a translation-invariant vacuum $\mathcal{V}$ of the 2d chiral algebra, where $X$ and $Y$ vevs are diagonal matrices with eigenvalues $(x_i,y_i)$ of multiplicity $N_i$.

The determinant operators we study are
\begin{equation}
{\cal D}(m;u;z) \equiv \det \left(m+ Z(u;z)\right)  = \int \dd \psi \dd \bar \psi \, e^{m\bar \psi \psi + \bar \psi Z(u;z) \psi} \, ,
\end{equation}
which can be expressed in terms of auxiliary (anti)fundamental fermions $\psi$ and $\bar\psi$. 

Holographically, the insertion of such a determinant represents the presence of a ``Giant Graviton" D-brane wrapping a 1-dimensional complex curve 
which approaches the boundary at a point $z$, along the line $x + u y + m=0$ \cite{Budzik:2021fyh}. 

In order to study correlation functions of multiple determinants, we follow the treatment of \cite{Jiang:2019xdz}, also implemented in \cite{Budzik:2021fyh, Chen:2019gsb, Berenstein:2022srd}.

We can use fermionization and normal ordering to express a correlator of determinants as
\begin{equation}
\prod_{a=1}^k {\cal D}(m_a;u_a;z_a) = \int \dd \psi \dd \bar \psi :e^{\sum_a \left[m_a \bar \psi_a \psi^a + \bar \psi_a Z(u_a;z_a) \psi^a \right]}: e^{- N^{-1}\sum_{a<b} \frac{u_a-u_b}{z_a-z_b} \bar \psi_a \psi^b \bar \psi_b \psi^a} \, .
\end{equation}
Then, we apply the Hubbard–Stratonovich transformation i.e. introduce an auxiliary bosonic $k\times k$ matrix $\rho$, with $\rho_a^a=m_a$:
\begin{equation}
\prod_{a=1}^k {\cal D}(m_a;u_a;z_a) = Z_\rho^{-1} \int \dd \psi \dd \bar \psi \, \dd'\rho :e^{\sum_{a,b}\rho^a_b \bar \psi_a \psi^b +\sum_a \bar \psi_a Z(u_a;z_a) \psi^a}: e^{N \sum_{a< b} \frac{z_a-z_b}{u_a-u_b} \rho^a_b \rho^b_a} \, ,
\end{equation}
where $\dd'\rho$ is the integration measure for the off-diagonal components of $\rho$ and $Z_\rho$ is a normalization factor. We evaluate the correlation function of chiral algebra fields in the vacuum $\mathcal{V}$:
\begin{align}
\expval{ : e^{\bar\psi_a Z(u_a;z_a) \psi^a } : }_{\mathcal{V}} &= e^{\bar\psi_a\psi^a \sum_{i=1}^n N_i(x_i+u_ay_i)}
\end{align}
The correlation function then becomes 
\begin{equation}
\Big\langle \prod_{a=1}^k {\cal D}(m_a;u_a;z_a) \Big\rangle_{\mathcal{V}} = Z_\rho^{-1} 
\int \dd'\rho \, e^{N \sum_{a< b} \frac{z_a-z_b}{u_a-u_b} \rho^a_b \rho^b_a} \prod_{i=1}^n \left[\det_{a,b} \rho^a_b + (x_i + u_a y_i)  \delta^a_b \right]^{N_i}  .
\end{equation}
Introducing diagonal $k \times k$ matrices $\zeta$, $\mu$ with entries $z_a$ and $u_a$ respectively, the large $N$ saddle equations become
\begin{equation} \label{eq:saddle1}
[\zeta, \rho] + \left[\mu, \sum_{i=1}^n \alpha_i \frac{1}{\rho + x_i + \mu y_i} \right] =0
\end{equation}
or 
\begin{equation} \label{eq:saddle2}
[\zeta, \rho] = \sum_{i=1}^n \alpha_i \frac{1}{\rho + x_i + \mu y_i}[\mu, \rho ] \frac{1}{\rho + x_i + \mu y_i} \, . 
\end{equation}

\subsection{The spectral curve}

In this section, we will map large $N$ saddles of determinant correlation functions studied above to complex curves in the dual geometry (\ref{eq:dualgeo}) using a spectral curve construction \cite{Budzik:2021fyh}.

For each saddle $\rho$ satisfying the equations (\ref{eq:saddle1})-(\ref{eq:saddle2}), we can define $k \times k$ matrices, functions of a spectral parameter $y$:
\begin{align}
X(y) &\equiv - \mu y - \rho \cr
Z_0(y) &\equiv \zeta - \sum_{i=1}^n \frac{\alpha_i}{y-y_i} \frac{1}{\rho + x_i + \mu y_i} \, .
\end{align}
The definition is such that 
\begin{equation}
[X(y),Z_0(y)]  = [\zeta, \rho] + \Big[\mu y + \rho,\sum_{i=1}^n \frac{\alpha_i}{y-y_i} \frac{1}{\rho + x_i + \mu y_i}\Big] =0 \, .
\end{equation}
We can look at simultaneous eigenvectors of $X(y)$ and $Z_0(y)$ as a function of $y$, with eigenvalues $x(y)$ and $z_{0}(y)$, away from $y=y_i$. 
This defines a holomorphic \textit{spectral curve} $\mathcal{S}_\rho$ in the $I_0$ patch of the expected dual geometry. 

We can also consider the matrix 
\begin{equation}
Z_\infty(y) = Z_0(y) - \sum_{i=1}^n \frac{\alpha_i}{y-y_i} \frac{1}{X(y)-x_i} =  \zeta - \sum_{i=1}^n \frac{\alpha_i}{y-y_i} \left[\frac{1}{\rho + x_i + \mu y_i}-\frac{1}{\rho + x_i + \mu y}\right] \, .
\end{equation}
We can also write it as 
\begin{equation}
Z_{\infty}(y) =  \zeta + \sum_{i=1}^n \alpha_i \frac{1}{\rho + x_i + \mu y_i}\mu \frac{1}{X(y)-x_i} \, .
\end{equation}
This matrix is regular at $y=y_i$ for $i=1,\dots,n$ and well-defined away from $x(y)=x_i$. It commutes with $X(y)$. 
We can look at simultaneous eigenvectors of $X(y)$ and $Z_\infty(y)$ as a function of $y$, with eigenvalues $x(y)$ and $z_{\infty}(y)$. 

If $y \neq y_i$, the matrix also commutes with $Z_0(y)$ and 
\begin{equation}
z_\infty(y) = z_0(y) - \sum_{i=1}^n \frac{\alpha_i}{y-y_i} \frac{1}{x(y)-x_i}\, .
\end{equation}
That means $x(y)$ and $z_\infty(y)$ extend the definition of the spectral curve $\mathcal{S}_\rho$ to the 
$I_\infty$ patch of the expected dual geometry (\ref{eq:dualgeo}). 

More generally, the collection of matrices  
\begin{equation}
Z_I(y) = Z_0(y) - \sum_{i\in I_x} \frac{\alpha_i}{y-y_i} \frac{1}{X(y)-x_i} \, ,
\end{equation}
commute with $X(y)$ and their eigenvectors $z_I(y)$ satisfy (\ref{eq:IIprime}) and therefore extend the spectral curve to all the patches of the expected dual geometry. 
The spectral curve ${\cal S}_\rho$ is thus a curve in the full geometry. 

We conjecture that the spectral curve ${\cal S}_\rho$ is the support of a B-model D-brane which is dual to the gauge theory saddle $\rho$. As a basic test, it approaches the boundary at $k$ locations, which at the leading order are at $x = u_a y$, $z_0=z_a$, $a=1,\dots,k$ in the $I_0$ patch. At sub-leading order, we find $x(y) + u_a y + \rho^a_a =0$. As $\rho^a_a=m_a$, this is the desired boundary condition of a brane dual to an insertion of a determinant $\mathcal{D}(u_a;z_a;m_a)$.

We can also compute the sub-leading behaviour of $z_0(y)$:
\begin{equation}
z_0(y) \sim z_a - y^{-1} \sum_{i=1}^n \alpha_i \left[\frac{1}{\rho + x_i + \mu y_i}\right]^a_a \, .
\end{equation}
Following the holographic dictionary \cite{Budzik:2021fyh}, we expect the coefficients on the right hand side: 
\begin{equation}
p^a = \sum_{i=1}^n \alpha_i \left[\frac{1}{\rho + x_i + \mu y_i}\right]^a_a
\end{equation}
to be the conjugate momentum to $m_a$ on the B-model side, i.e. the derivative of the D-brane action with respect to $m_a$. 
At the same time, we recognize $p^a = \frac{\partial S}{\partial m_a}$ for the semiclassical action $S$ at the gauge theory saddle. 
This insures that the action for this B-model D-brane matches the semiclassical action for the gauge theory saddle, up to $m$-independent terms. 

\subsection{Restoring $SL(2)_R$ invariance}
Our formalism in this section is not manifestly covariant under $SL(2)_R$. If we define
\begin{equation}
Z_{v^{-1}}(y) = Z_0(y)  + \sum_{i=1}^n \frac{\alpha_i}{(y - y_i)}\frac{1}{X(y)+ v y-x_i- v y_i} \, ,
\end{equation}
which can be rewritten as
\begin{equation}
Z_{v^{-1}}(y) =\zeta + \sum_{i=1}^n \alpha_i \frac{1}{\rho + x_i + \mu y_i}(\mu-v) \frac{1}{X(y)+ v y-x_i- v y_i} \, ,
\end{equation}
we can see that when acting on an eigenvector $V$ of $X(y)$ and $Z(y)$ with eigenvalues $x$ and $z_0$ we have 
\begin{align}
z_{v^{-1}}V &=\zeta V + \sum_{i=1}^n \alpha_i  \frac{1}{x+ v y-x_i- v y_i} \frac{1}{(\mu-v)^{-1} \rho + (\mu-v)^{-1} (x_i + v y_i) +  y_i}V \\
&\equiv Z_{v^{-1}}(x+v y) V
\end{align}
and
\begin{equation}
 y V =-(\mu-v)^{-1} \rho V- (x+ v y) (\mu-v)^{-1} V\equiv Y(x+v y) V \, .
\end{equation}
That means $V$ is a simultaneous eigenvector for the matrices $Y(x+v y)$ and $Z_{v^{-1}}(x+v y)$, with eigenvalues $z_{v^{-1}}(x+v y)$ and $y(x+v y)$. 
These are built just as $Z_0(y)$ and $X(y)$, with the matrix $(\mu-v)^{-1} \rho$ playing the role of $\rho$.

This shows that the construction of the spectral curve ${\cal S}_\rho$ is $SL(2)_R$-covariant.  

\subsection{Single-trace insertion and bulk-to-boundary propagator}
A single-trace insertion in the correlation function of determinants can also be evaluated in the large $N$ limit \cite{Jiang:2019xdz}. Without loss of generality, we can look at the insertion of $\Tr Y^n(w)$, or better a generating function 
\begin{equation}
N \mathrm{Tr} \log \left( \hat y- Y(w) \right) =  N \mathrm{Tr} \log \hat y - \sum_{n=1}^\infty \frac{1}{n \hat y^n} N \mathrm{Tr}Y(w)^n \, . \label{eq:resolvent}
\end{equation}
The main effect of vacuum vevs on Feynman diagrams is that $Y$ fields can either be contracted with a propagator or replaced by their vev. The dominant Feynman diagrams have the topology of a disk, with the auxiliary fermions running along the boundary.

After some combinatorial manipulations analogous to these in \cite{Budzik:2021fyh}, the correlation function is the sum of the classical vev $N \mathrm{Tr} \log \left(\hat y- Y_\infty \right)$ and the leading quantum correction 
\begin{equation}
-N \mathrm{Tr}_{k\times k}  \log \left(1- \frac{1}{w-\zeta} \sum_{i=1}^n \frac{1}{\hat y-y_i} \frac{\alpha_i}{\rho + x_i + \mu y_i} \right) \, .
\end{equation}
This can be identified with 
\begin{equation}
-N \mathrm{Tr}_{k\times k}  \log \frac{w-Z_0(\hat y)}{w-\zeta} = N \sum_{a=1}^k \log (w-z_a) -N \log \det \left(w-Z_0(\hat y)\right) \, .
\end{equation}
The right hand side can be computed from a contour integral 
\begin{equation} \label{eq:fin}
\oint \log (\hat y-y') \partial_{y'}\log \det \left(w-Z_0(\hat y)\right) = \sum_{y'_*(w)} \log(\hat y-y'_*(w)) 
\end{equation}
and evaluated as a sum over the intersection points $y'_*$ of the spectral curve and $z_0=w$. 

We can now compare this with a B-model calculation of the leading Witten diagram: the integral of a boundary-to-bulk propagator $\partial^{-1} \alpha_w$ 
on the conjectural D-brane world-volume ${\cal S}_\rho$. We expect to have a representative for the bulk-to-boundary propagator of the KS theory for $\mathrm{Tr} Y^n(w)$ of the form
\begin{equation}
\partial^{-1} \alpha_{n;w}= y^n \delta_{\Delta_{z_0=w}}  \, .
\end{equation}
Thus the bulk-to-boundary propagator for the insertion (\ref{eq:resolvent}) is
\begin{equation}
\partial^{-1} \alpha_{w}[\hat y]= \log(\hat y - y) \delta_{\Delta_{z_0=w}}  \, .
\end{equation}
It is straightforward to integrate $\partial^{-1} \alpha_{w}[\hat y]$ on the spectral curve: it only receives contributions from the intersections between the spectral curve and $z_0=w$ and reproduces (\ref{eq:fin}).

\section{Acknowledgments} 
It is a pleasure to thank Tomoyuki Arakawa, Kevin Costello, Jaume Gomis, Ji Hoon Lee and Aiden Suter for useful conversations. This research was supported in part by a grant from the Krembil Foundation. DG is supported by the NSERC Discovery Grant program and by the Perimeter Institute for Theoretical Physics. Research at Perimeter Institute is supported in part by the Government of Canada through the Department of Innovation, Science and Economic Development Canada and by the Province of Ontario through the Ministry of Colleges and Universities.
\bibliographystyle{JHEP}

\bibliography{mono}

\end{document}